\documentclass{article}
\usepackage{spconf,amsmath,graphicx}
\usepackage{multirow}
\usepackage{multicol}
\usepackage{listings}

\usepackage{url}
\usepackage{todonotes} 

\title{Improving sequence-to-sequence speech recognition training with on-the-fly data augmentation}
\name{Thai-Son Nguyen$^1$, Sebastian St\"uker$^1$, Jan Niehues$^2$, Alex Waibel$^1$}
\address{$^1$ Institute for Anthropomatics and Robotics, Karlsruhe Institute of Technology \\
$^2$Department of Data Science and Knowledge Engineering (DKE), Maastricht University}
\begin{document}
%
\maketitle
\begin{abstract}
Sequence-to-Sequence (S2S) models recently started to show state-of-the-art performance for automatic speech recognition (ASR). With these large and deep models overfitting remains the largest problem, outweighing performance improvements that can be obtained from better architectures. One solution to the overfitting problem is increasing the amount of available training data and the variety exhibited by the training data with the help of data augmentation. In this paper we examine the influence of three data augmentation methods on the performance of two S2S model architectures. One of the data augmentation method comes from literature, while two other methods are our own development – a time perturbation in the frequency domain and sub-sequence sampling.
Our experiments on Switchboard and Fisher data show state-of-the-art performance for S2S models that are trained solely on the speech training data and do not use additional text data.
\end{abstract}
\begin{keywords}
Sequence-to-sequence, Self-attention, Data Augmentation, Speed Perturbation, Sub-sequence
\end{keywords}

\vspace{-0.1cm}
\section{Introduction}
\label{sec:intro}

In automatic speech recognition (ASR), data augmentation has been used for producing additional training data in order to increase the quality of the training data, i.e. their amount and variety. This then improves the robustness of the models and avoids overfitting. As in \cite{kanda2013elastic,ragni2014data}, both unsupervised and artificial training data has been augmented to improve model training in low-resource conditions. The addition of training data with perturbation of the vocal tract length \cite{jaitly2013vocal} or audio speed \cite{ko2015audio} helps ASR models to be robust to speaker variations. Simulated far-field speech \cite{ko2017study} and noisy speech \cite{hannun2014deep} have been used to supplement clean close-talk training data.

Sequence-to-sequence attention-based models \cite{chorowski2015attention,chan2015listen} were introduced as a promising approach for end-to-end speech recognition. Several advances \cite{chiu2018state,zeyer2018improved,weng2018improving} have been proposed for improving the performance of S2S models. While many works focus on designing better network architectures, the authors in \cite{park2019specaugment} have recently pointed out that overfitting is the most critical issue when training their sequence-to-sequence model on popular benchmarks. By proposing a data augmentation method together with a long training schedule to reduce overfitting, they have achieved a large gain in performance superior to many modifications in network architecture.


To date, there have been different sequence-to-sequence encoder-decoder models \cite{park2019specaugment,pham2019very} reporting superior performance over the HMM hybrid models on standard ASR benchmarks. While \cite{park2019specaugment} uses Long Short-Term Memory (LSTM) networks, for both encoder and decoder, \cite{pham2019very} employs self-attention layers to construct the whole S2S network.

In this paper, we investigate three on-the-fly data augmentation methods for S2S speech recognition, two of which are proposed in this work and the last was recently discovered~\cite{park2019specaugment}. We contrast and analyze both LSTM-based and self-attention S2S models that were trained with the proposed augmentation methods by performing experiments on the Switchboard (SWB) and Fisher telephone conversations task. We found that not only the two models behave differently with the augmentation methods, but also the combination of different augmentation methods and network architectures can significantly reduce word error rate (WER). Our final S2S model achieved a WER of 5.2\% on the SWB test set and 10.2\% WER on the Callhome (CH) test set. This is already on par with human performance. We made our source code available as open source\footnote{The source code is available at \textit{https://github.com/thaisonngn/pynn}}, as well as the model checkpoints of the experiments in this paper.

\vspace{-0.2cm}
\section{Data Augmentation}
\label{sec:da}
We investigated three data augmentation methods for sequence-to-sequence encoder-decoder models. The first two modify the input sequences from different inspirations and aim to improve the generalization of the log-mel spectrogram encoder. The third approach improves the decoder by adding sub-samples of target sequences. All of the proposed methods are computationally cheap and can be performed on-the-fly and can be optimized together with the model.

\vspace{-0.2cm}
\subsection{Dynamic Time Stretching}
\label{ssec:da_timestretch}
Many successful S2S models adopt log-mel frequency features as input. In the frequency domain, one major difficulty for the recognition models is to recognize temporal patterns which occur with varying duration. To make the models more robust to temporal variations, the addition of audio data with speed perturbation in the time domain such as in \cite{ko2015audio} has been shown to be effective. In contrast, in our work we manipulate directly the time series of the frequency vectors which are the features of our S2S models, in order to achieve the effect of speed perturbation. Specifically, given a sequence of consecutive feature vectors $seq$, we stretch every window of $w$ feature vectors by a factor of $s$ obtained from an uniform distribution of range $[low, high]$, resulting in a new window of size $w*s$. There are different approaches to perform window stretching, in this work we adopt \textit{nearest-neighbor interpolation} for its speed, as it is fast enough to augment many speech utterances on a CPU while model training for other utterances is being performed on a GPU. The dynamic time stretching algorithm is implemented by the following python code:
\begin{lstlisting}[language=Python,basicstyle=\small]
def time_stretch(seq,w,low=0.8,high=1.25):
    ids = None; time_len = len(seq)
    for i in range(time_len // w + 1):
        s = random.uniform(low, high)
        e = min(time_len, w*(i+1))          
        r = numpy.arange(w*i, e-1, s)
        r = numpy.round(r).astype(int)
       ids = numpy.concatenate((ids, r))
    return seq[ids]
\end{lstlisting}

\vspace{-0.4cm}
\subsection{SpecAugment}
\label{ssec:da_specaugment}
Recently \cite{park2019specaugment} found that LSTM-based S2S models tend to overfit easily to the training data, even when regularization methods such as Dropout \cite{srivastava2014dropout} are applied. Inspired by the data augmentation from computer vision, \cite{park2019specaugment} proposed to deform the spectrogram input with three cheap operations such as time warping, frequency and time masking before feeding it to their sequence-to-sequence models. Time warping shifts a random point in the spectrogram input with a random distance, while frequency and time masking apply zero masks to some consecutive lines in both the frequency and the time dimensions. In this work, we study the two most effective operations which are the frequency and time masking. Experimenting on the same dataset, we benefit from optimized configurations in \cite{park2019specaugment}. Specifically, we consider $T \in [1,2,3]$ -- the number of times that both, frequency and time masking, are applied. For each time, $f$ consecutive frequency channels and $t$ consecutive time steps are masked where $f$ and $t$ are randomly chosen from $[0, 70]$ and $[0, 7]$. When $T = 2$, we obtain a similar setting for 40 log-mel features as the SWB mild (SM) configuration in \cite{park2019specaugment}. We experimentally find $T$ for different model architectures in our experiments.

\vspace{-0.4cm}
\subsection{Sub-sequence Sampling}
\label{ssec:da_subsequence}
\begin{figure}[t]
	\centering
	\includegraphics[width=0.90\linewidth]{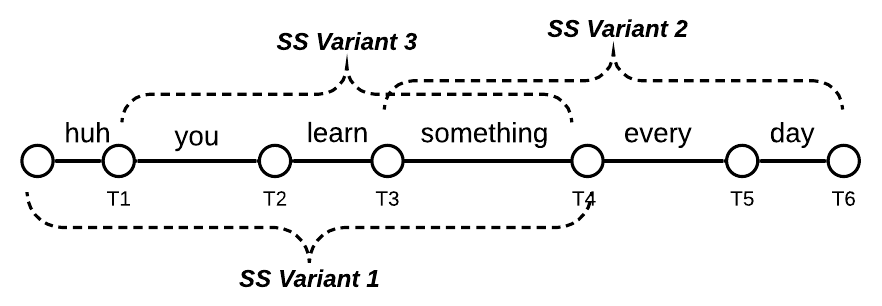}
	\vspace{-0.5cm}
	\caption{Sub-sequence Sampling.}
	\label{fig:ss}
    \vspace{-0.5cm}
\end{figure}

Different from other S2S problems, the input-output of speech recognition models are the sequences of speech feature vectors and label transcripts which are monotonically aligned. The alignment can be also estimated automatically via the traditional force-alignment process. Taking advantage of this property, we experiment with the ability to sub-sample training utterances to have more variants of target sequences. Since the approach of generating sub-sequences with arbitrary lengths does not work, we propose a constraint sampling depicted in Figure~\ref{fig:ss}. Basically, given an utterance, we allow three different variants of sub-sequences with equal distributions. The first and second variants constraint sub-sequences to having either the same start or end as the original sequence while the third variant needs to have their start and end point within the utterance. All sub-sequences need to have at least half the size of the original sequence. During training, we randomly select a training sample with probability $alpha$ and replace it with one of the sampled sub-sequence variants. We also allow $static$ mode in which only one fixed instance of sub-sequence per utterance per variant is generated. This mode is equivalent to statically adding three sets of sub-sequences to the original training set.
\vspace{-0.4cm}

\vspace{-0.0cm}
\section{Model}
\label{sec:models}
We use two different S2S models to investigate the on-the-fly data augmentation methods proposed in Section \ref{sec:da}. In the first model, we use LSTMs and a new approach for building the decoder network. For the second model, we follow the work in \cite{pham2019very} to replace LSTMs with deep self-attention layers in both the encoder and decoder.

\vspace{-0.3cm}
\subsection{LSTM-based S2S}
\label{ssec:models_lstm}
Before the LSTM layers in the encoder, we place a two-layer Convolutional Neural Network (CNN) with 32 channels and a time stride of two to down-sample the input spectrogram by a factor of four. In the decoder, we adopt two layers of unidirectional LSTMs as language modeling for the sequence of sub-word units and the approach of Scaled Dot-Product (SDP) Attention \cite{vaswani2017attention} to generate context vectors from the hidden states of the two LSTM networks. Specifically, our implementation for LSTM-based S2S works as follows:
\vspace{-0.1cm}
\begin{eqnarray*}
	enc = LSTM(CNN(spectrogram)) \\
	emb = Embedding(subwords) \\
	dec = LSTM(emb) \\
	context = SDPAttention(dec, enc, enc) \\	
	y = Distribution(context + dec)
\end{eqnarray*}
\vspace{-0.2cm}
Different from previous works \cite{chan2016listen,chiu2018state,zeyer2018improved,weng2018improving}, we adopt a simpler recurrent function in the decoder (i.e. without Input-feeding \cite{weng2018improving}), and a more complicated attention module. The adopted attention function learns an additional linear transformation for each input parameter (known as query, key and value) and use the multi-head mechanism together with Dropout and LayerNorm for efficiently learning content-based attention~\cite{vaswani2017attention}. In fact, the implementation of the attention function is shared with the deep self-attention network from Section \ref{ssec:models_transformer}. In addition to that, we share the parameters between $Embedding$ and $Distribution$ to improve the word embedding. Because this implementation does not require us to customize LSTM cells (which is needed by Input-feeding), we can achieve high parallelization~\footnote{Highly optimized LSTM implementation offered by cuDNN library} to speed up training.

\vspace{-0.2cm}
\subsection{Self-Attention S2S}
\label{ssec:models_transformer}
We follow \cite{pham2019very} to build an encoder-decoder model with deep self-attention layers. Specifically, we use many stochastic self-attention layers (e.g., 36 and 12) for the encoder and the decoder for better generalization of the deep architecture. Instead of using a CNN for down-sampling the input spectrogram, we stack four consecutive feature vectors after applying the augmentation methods. Compared to \cite{pham2019very}, we use BPE sub-word units instead of characters for target sequences. For more details refer to \cite{pham2019very}.

\vspace{-0.3cm}
\section{Experimental Setup}
\label{sec:setups}
\vspace{-0.2cm}
Our experiments were conducted the Switchboard (300 hours) and the Fisher+Switchboard (2000h) corpora. The Hub5'00 evaluation data was used as the test set. For input features, we use 40 dimensional log-mel filterbanks normalized per conversation. For labels, SentencePiece was used for generating 4,000 BPE sub-word units from all the transcripts.
We use Adam \cite{kingma2014adam} with an adaptive learning rate schedule defined by \textit{(lr, warm-up, decay)} in which the learning rate \textit{lr} increases for the first \textit{warm-up} steps and then decreases linearly. We adopted the approach in \cite{pham2019very} for the exact calculation of the learning rate at every step. In addition to that, we further decay the learning rate exponentially with a factor of 0.8 after every \textit{decay} step. We save the model parameters of 5 best epochs according to the cross-validation sets and average them at the end.

\vspace{-0.3cm}
\section{Results}
\label{sec:results}

\vspace{-0.2cm}
\subsection{Baseline Performance}
\label{ssec:result_baseline}
Using the SWB material and an unique label set of 4k sub-words, we trained both of the proposed S2S models for 50 epochs. We adopt a mini-batch size of 8,000 label tokens which contains about 350 utterances. In our experiments, the LSTM-based models tend to overfit after 12k updates (i.e. perplexity increases on the cross-validation set) while the self-attention models converge slower and saturate at 40k updates. We were able to increase the size of the LSTM-based as well as the depth of the self-attention models for performance improvement. We stop at six layers of 1,024 units for the encoder of the LSTM-based and 36-12 encoder-decoder layers of self-attention models, and then use them as baselines for further experiments. Table~\ref{tab:baseline} shows the WER of the baselines. We also include the results of the baseline models when trained on the speed-perturbed dataset~\cite{ko2015audio}.
\begin{table}[t]
	\setlength{\tabcolsep}{6pt}
	\centering
	\begin{tabular}{|c|c|c|c|c|}
		\hline
		Model & Size & SWB & CH & Hub5'00 \\
		\hline \hline
		\multirow{3}{*}{LSTM} & 4x512 & 12.9 & 24.1 & 18.5 \\
		& 6x1024 & 12.1 & 22.7 & 17.4 \\
		& 6x1024 \textit{(SP)} & 10.7 & 20.5 & 15.6 \\
		\hline
		\multirow{3}{*}{Transformer} & 8x4 & 13.2 & 24.7 & 19.0 \\
		& 36x12 & 11.1 & 21.1 & 16.1 \\
		& 36x12 \textit{(SP)} & 10.2 & 19.4 & 14.8 \\
		\hline
	\end{tabular}
	\caption{\label{tab:baseline} {Baseline models using Switchboard 300h.}}
	\vspace{-0.2cm}
\end{table}

\vspace{-0.3cm}
\subsection{Time Stretching and SpecAugment}
\label{ssec:result_ts_sa}
\begin{table}[t]
	\setlength{\tabcolsep}{6pt}
	\centering
	\begin{tabular}{|c|c|c|c|}
		\hline
		\multirow{1}{*}{TimeStretch} & \multirow{1}{*}{SpecAugment} &
		\multirow{1}{*}{LSTM} & \multirow{1}{*}{Transformer} \\
		\textit{w} & \textit{T} & Hub5'00 & Hub5'00 \\
		\hline
		50   & - & 16.1 & 15.5 \\		
		100  & - & 15.9 & 14.9 \\
		200  & - & 16.0 & 14.9 \\
	$\infty$ & - & 16.1 & 15.0 \\
		 - & 1 & 14.7 & 14.3 \\
		 - & 2 & 14.1 & 14.5 \\
		 - & 3 & 14.3 & 14.4 \\
		100 & 1 & 14.2 & 13.8 \\		  
	$\infty$ & 1 & 13.9 & 13.6 \\
		100 & 2 & 13.7 & 13.9 \\		  
	$\infty$ & 2 & 13.6 & 13.7 \\
		\hline
	\end{tabular}
	\caption{\label{tab:ts_sa} {The performance of the models trained with \textit{TimeStretch} and \textit{SpecAugment} augmentation.}}
	\vspace{-0.4cm}
\end{table}
Both \textit{Time Stretching} and \textit{SpecAugment} are augmentation methods which modify the input sequences aiming to improve the generalization of the encoder network. We trained several models for evaluating the effects of these methods individually as well as the combinations as shown in Table~\ref{tab:ts_sa}.

For \textit{Time Stretching}, WER slightly changed when using different window sizes. However the 8.6\% and 12.4\% rel. improvement over the baseline performance of the LSTM-based and self-attention models clearly shows its effectiveness. With a window size of 100ms, the models can nearly achieve the performance of the static speed perturbation augmentation.

As shown in \cite{park2019specaugment}, \textit{SpecAugment} is a very effective method for avoiding overfitting on the LAS model. Using this method, we can also achieve a large WER improvement for our LSTM-based models. However, our observation is slightly different from \cite{park2019specaugment}, as \textit{SpecAugment} slows down the convergence of the training on the training set and significantly reduces the loss on the validation set (as for \textit{Time Stretching}) but does not change from overfitting to underfitting. The losses of the final model and the baseline model computed on the original training set are similar. 

\textit{SpecAugment} is also effective for our self-attention models. However, the improvements are not as large as for the LSTM-based models. This might be due to the self-attention models not suffering from the overfitting problem as much as the LSTM-based models. It is worth noting that for the self-attention models, we use not only Dropout but also \textit{Stochastic Layer}~\cite{pham2019very} to prevent overfitting. When tuning $T$ for both models, we observed different behaviours. The LSTM-based models work best when $T=2$, but for self-attention, different values of $T$ produce quite similar results. This might be due to the fact that the self-attention encoder has direct connections to all input elements of different time steps while the LSTM encoder uses recurrent connections.

When combining two augmentation methods within a single training (i.e. applying \textit{Time Stretching} first and then \textit{SpecAugment} for input sequences), we can achieve further improvements for both models. This result indicates that both methods help the models to generalize across different aspects and can supplement each other. We keep using the optimized settings ($T=2$ and $w=\infty$ for LSTM-based and $T=1$ for self-attention) for the rest of the experiments.
\begin{table}[t]
	\setlength{\tabcolsep}{5pt}
	\centering
	\begin{tabular}{|c|c|c|c|}
		\hline
		\multirow{1}{*}{Sub-sequence} & \multirow{1}{*}{SpecAugment} &
		\multirow{1}{*}{LSTM} & \multirow{1}{*}{Transformer} \\
		\textit{alpha} & \& TimeStretch & Hub5'00 & Hub5'00 \\
		\hline
		0.3 & N & 18.6 & 15.6 \\
		0.5 & N & 18.6 & 15.4 \\		
		0.7 & N & 18.8 & 15.3 \\
		0.7 (static) & N & 15.4 & 15.1 \\
		0.7 & Y & 13.5 & 13.4 \\
		0.7 (static) & Y & 13.0 & 13.2 \\
		\hline
	\end{tabular}
	\caption{\label{tab:subseq} {The performance of the models trained with \textit{Sub-sequence} augmentation.}}
	\vspace{-0.6cm}
\end{table}

\vspace{-0.4cm}
\subsection{Combining with Sub-sequence}
\label{ssec:results_ss}
Table~\ref{tab:subseq} presents the models' performance when we applied \textit{Sub-sequence} augmentation with different $alpha$ values. We observe contrary results for different models: improving the self-attention but downgrading the performance of the LSTM-based models. These observations are indeed consistent with the overfitting problems observed with the two models. The LSTM-based models even overfit more quickly to the dataset with sub-sequence samples while self-attention models do not, so that they can benefit from \textit{Sub-sequence}. However, when using a static set of sub-sequences, we obtained clear improvement for LSTM-based models but had comparable performance for self-attention models. This reveals an interesting observation for the differences between self-attention and LSTM when interpreting them as language models in the decoder. The static approach is also better when combined with other augmentation methods.
\begin{table}[t]
	\setlength{\tabcolsep}{7pt}
	\centering
	\begin{tabular}{|l|c|c|c|}
		\hline
		Model & LM & SWB & CH \\
		\hline \hline
		\multicolumn{4}{|c|}{\textit{300h Switchboard}} \\
		\hline
		Zeyer et al. 2018 \cite{zeyer2018improved} & LSTM & 8.3 & 17.3 \\
		Yu et al. 2018 \cite{zeyer2018improved} & LSTM & 11.4 & 20.8 \\
		Pham et al. 2019 \cite{pham2019very} & - & 9.9 & 17.7 \\
		Park et al. 2019 \cite{park2019specaugment} & LSTM & 7.1 & 14.0 \\
		Kurata et al. 2019 \cite{kurata2019guiding} & - & 11.7 & 20.2 \\
		\hline
		\textit{LSTM-based} & - & 8.8 & 17.2 \\
		\textit{Transformer} & - & 9.0 & 17.5 \\
		\textit{ensemble} & - & 7.5 & 15.3 \\
		\hline
		\multicolumn{4}{|c|}{\textit{2000h Switchboard+Fisher}} \\
		\hline
		Povey et al. 2016 \cite{povey2016purely} & n-gram & 8.5 & 15.3 \\
		Saon et al. 2017 \cite{saon2017english} & LSTM & 5.5 & 10.3 \\
		Han et al. 2018 \cite{han2017capio} & LSTM & 5.0 & 9.1 \\
		Weng et al. 2018 \cite{weng2018improving} & - & 8.3 & 15.5 \\
		Audhkhasi et al. 2018 \cite{audhkhasi2018building} & - & 8.8 & 13.9 \\
		\hline
		\textit{LSTM-based (no augment.)} & - & 7.2 & 13.9 \\		
		\textit{Transformer (no augment.)} & - & 7.3 & 13.5 \\
		\textit{LSTM-based} & - & 5.5 & 11.4 \\
		\textit{Transformer} & - &6.2 & 11.9 \\
		\textit{ensemble} & - & 5.2 & 10.2 \\
		\hline
	\end{tabular}
	\caption{\label{tab:result_final} {Final performance on Switchboard 300h and Fisher 2000h training sets.}}
	\vspace{-0.6cm}
\end{table}

\vspace{-0.2cm}
\subsection{Performance on Full Training Set}
We report the final performance of our models trained on the 2,000h in Table~\ref{tab:result_final}. Slightly different from 300h, we used a larger mini-batch size of 12k tokens and do not use the exponential decay of the learning rate. We also increased the model size by a factor of 1.5 while keeping the same depth. We need 7 hours to finish one epoch for the LSTM-based models, 3 hours for the self-attention models. With the bigger training set, the LSTM-based models saturate after 100k updates while the self-attention models need 250k updates. Even with the large increase in training samples, the proposed augmentation is still effective since we observe clear gaps between the models with and without augmentation. For the final performance, we found that the ensemble of the LSTM-based and self-attention models are very efficient for the reduction of WER. Our best performance on this benchmark is competitive compared to the best performance reported in the literature so far, and it is notable that we did not employ any additional text data, e.g., for language modeling.

\vspace{-0.4cm}
\section{Conclusion}
\label{sec:conclusion}
We have shown the improvements obtained from three data augmentation techniques when applied to two different architectures of S2S modeling. By utilizing these techniques we were able to achieve state of the art performance on the Switchboard and CallHome test sets when not utilizing additional language models. Future work will evaluate different algorithms for stretching the window of feature vectors and different strategies for sub-sampling.


\bibliographystyle{IEEEbib}
\bibliography{refs}

\end{document}